\documentclass[prd,reprint,no title page,twoside,onecolumn,showpacs,
               nofootinbib,longbibliography,amsmath,amssymb,
               floatfix]{revtex4-1}

\usepackage{graphicx}   % Include figure files
\usepackage{bm}         % bold math

\mathchardef\ordinarycolon\mathcode`\:                     %
\def\vcentcolon{\mathrel{\mathop\ordinarycolon}}           % Better versions
\providecommand*\coloneqq{{\;\vcentcolon\mkern-0.8mu=\;}}  % of := and =:

\begin{document}

\enlargethispage {0.5in}

\title{Gravity Inside a Nonrotating, Homogeneous, Spherical Body}

\author{Homer G.~Ellis}
\affiliation{Department of Mathematics, University of Colorado at Boulder,
Boulder, Colorado 80309}

\date{October 11, 2012}

\begin{abstract}
Schwarzschild's `interior solution' is a space-time metric that satisfies 
Einstein's gravitational field equations with a source term that Einstein
created on the basis of an unjustified identification of the conceptually
distinct notions of the passive gravitational mass of matter and the active
gravitational mass of matter.  Giving up that identification allows deriving
from a variational principle new and better (because logically obtained) field
equations that more faithfully extend the Poisson equation for Newton's gravity
than do Einstein's, with an active gravitational mass density providing the
source term.  Solving these equations for a nonrotating spherical ball of
matter with uniform density produces a new, improved interior metric matched at
the surface of the ball to Schwarzschild's `exterior' solution metric.  This
new metric can then be used to address questions about the flight times of
photons and neutrinos through such a ball of matter.
\end{abstract}


\maketitle

In 1916 Karl Schwarzschild published his static, spherically symmetric solution
of Einstein's vacuum field equations~\cite{schw1} (the `Schwarzschild
blackhole' solution), taken to represent, in its entirety, the gravitational
field of a point particle and, when restricted to values of the radial
coordinate greater than $R$, the gravitational field external to a massive
spherically symmetric body of radius $R$ (the `Schwarzschild exterior
solution').  Later in the same year he published a solution of Einstein's
nonvacuum field equations, to represent the gravitational field inside a
``homogeneous sphere\ldots consisting of incompressible fluid''~\cite{schw2},
the solution commonly referred to as the `Schwarzschild interior solution'.

The metric of Schwarzschild's interior solution takes the proper-time form
\begin{equation}
d\tau^2 =
 \left[ \frac32 \left(1 - \frac{2 m}{R}\right)^\frac12
       -\frac12 \left(1 - \frac{2 m}{R} \frac{r^2}{R^2}\right)^\frac12
 \right]^2 dt^2
  - \frac{1}{c^2} \left(1 - \frac{2 m}{R} \frac{r^2}{R^2}\right)^{\!-1} \! dr^2
  - \frac{1}{c^2} \, r^2 \, d\Omega^2 \, ,
\label{eqn1}
\end{equation}
in a coordinate system (not used by Schwarzschild) chosen to facilitate at
$r = R$ as smooth a matching as possible to the exterior solution
\begin{equation}
d\tau^2 =
 \left(1 - \frac{2 m}{r}\right) dt^2
  - \frac{1}{c^2} \left(1 - \frac{2 m}{r}\right)^{\!-1} \! dr^2 - \frac{1}{c^2} \, r^2 \, d\Omega^2 \, .
\label{eqn2}
\end{equation}
Einstein's nonvacuum field equations that Schwarzschild's interior solution
satisfies are
\begin{equation}
{\bm R}^{\alpha \beta} - {\textstyle\frac12} {\bm R} g^{\alpha \beta}
 = \frac{8 \pi \kappa}{c^2} T^{\alpha \beta} \, ,
\label{eqn3}
\end{equation} 
where
$T^{\alpha \beta} = (\rho + p/c^2) u^\alpha u^\beta - (p/c^2) g^{\alpha \beta}$ (obtained from Einstein's $T^{\alpha \beta} =
\rho u^\alpha u^\beta - (p/c^2) g^{\alpha \beta}$ by his redefinition
$\rho \to \rho + p/c^2$~(\cite{eins1},\S 16)).
Here $\kappa$ is Newton's gravitational constant, $p$ is the pressure of the
``incompressible fluid'', $u$ is its velocity field, and $\rho$ is its
``density'', a quantity described imprecisely by Einstein as the ``density of
matter'' (of the fluid, in Schwarzschild's case).  

I have shown elsewhere~\cite{elli1} that this use of $\rho$ and $p$ in
Eq.~(\ref{eqn3}) constitutes an unjustified identification of
inertial-{\it passive} mass density ($\rho$) and kinetic energy ($p$), which
have no proper place in a source term for gravitational field equations, with
{\it active} gravitational mass density $\mu$, which does belong there.  To
correct this error I have derived new and better (because logically obtained)
field equations from the variational principle
\begin{equation}
\delta \! \int ({\bm R}
 - \frac{8 \pi \kappa}{c^2} \mu
 + 2 \, \phi^{.\gamma} \phi_{.\gamma}) \, |g|^{\frac12} \, d^4\!x = 0 \, ,
\label{eqn4}
\end{equation}
in which $\phi$ is an `auxiliary' scalar field introduced to allow a larger
class of solutions of the field equations than would be available without it.
Without the $\phi$ term this variational principle is the most straightforward
generalization to the general relativity setting of the variational principle
$\delta \! \int (|\nabla V|^2 + 8 \pi \kappa \mu V) \, d^3\!x = 0$ that
produces the Poisson equation $\nabla^2 V = 4 \pi \kappa \mu$ for the newtonian
gravitational potential $V$.  The field equations that result from
Eq.~(\ref{eqn4}) when {\it only} the metric is varied are
\begin{equation}
{\bm R}_{\alpha \beta} - {\textstyle\frac12} {\bm R} \, g_{\alpha \beta}
  = -\frac{4 \pi \kappa}{c^2} \mu \, g_{\alpha \beta}
     - 2 \, (\phi_{.\alpha} \phi_{.\beta}
             - {\textstyle\frac12} \phi^{.\gamma} \phi_{.\gamma}
                \, g_{\alpha \beta}) \, ,
\label{eqn5}
\end{equation}
and
\begin{equation}
2 \, (\square \phi) \phi_{.\alpha}
  \coloneqq 2 \, \phi^{.\gamma}\!{}_{:\gamma} \phi_{.\alpha}
  = -\frac{4 \pi \kappa}{c^2} \mu_{.\alpha} \, ,
\label{eqn6}
\end{equation}
the latter of which follows from the vanishing of the divergence of
${\bm R}_{\alpha \beta} - {\textstyle\frac12} {\bm R} \, g_{\alpha \beta}$
in the former and is equivalent to $\square \phi = 0$ when $\mu$ is a constant.
Equivalent to Eq.~(\ref{eqn5}) is
\begin{equation}
{\bm R}_{\alpha \beta}
  = \frac{4 \pi \kappa}{c^2} \mu \, g_{\alpha \beta}
    - 2 \, \phi_{.\alpha} \phi_{.\beta} \, .
\label{eqn7}
\end{equation}

\begin{figure}
\includegraphics[width=5.0in,height=2.25in]{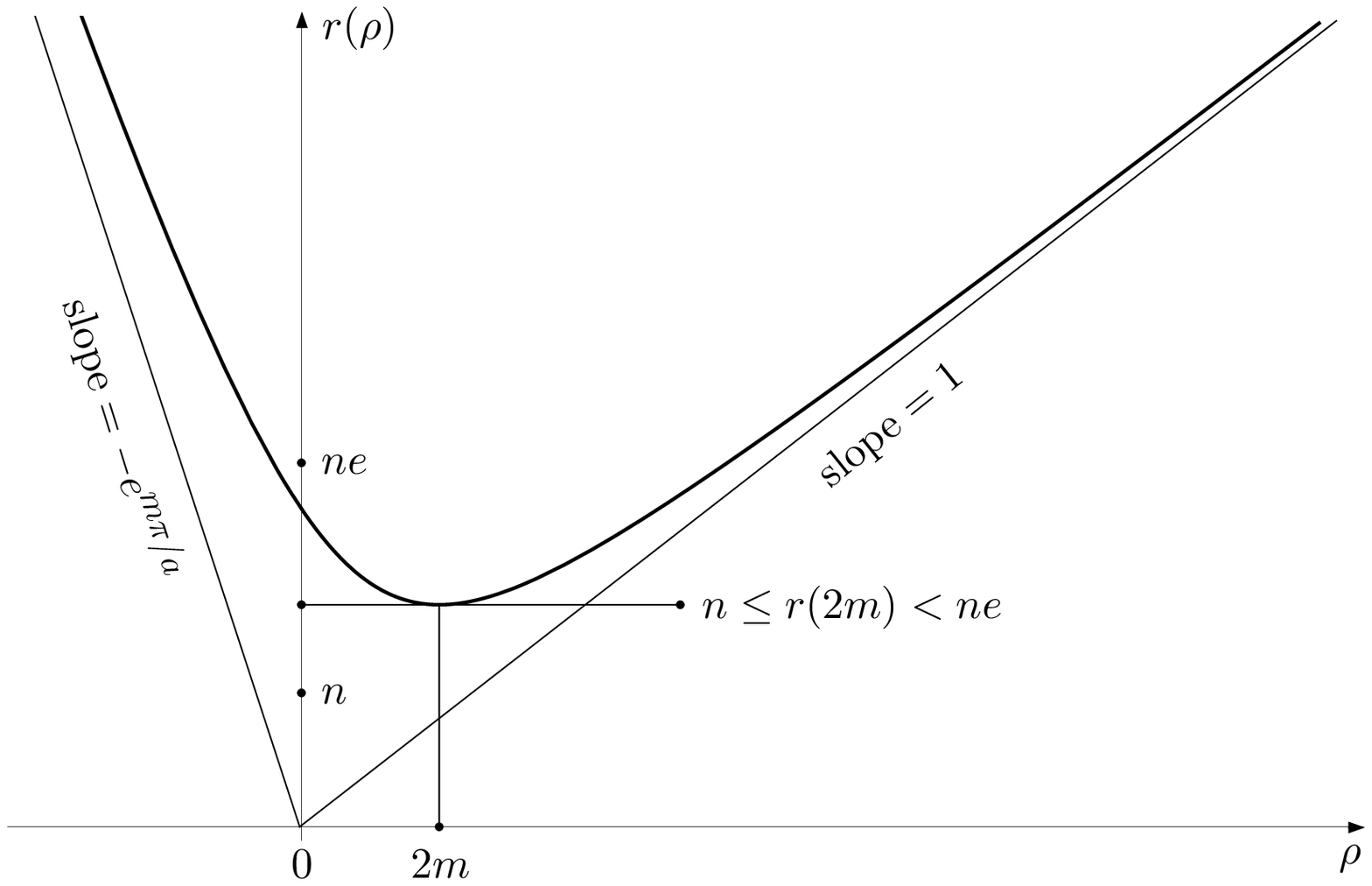}
\vskip 15pt
\includegraphics[width=5.0in,height=2.25in]{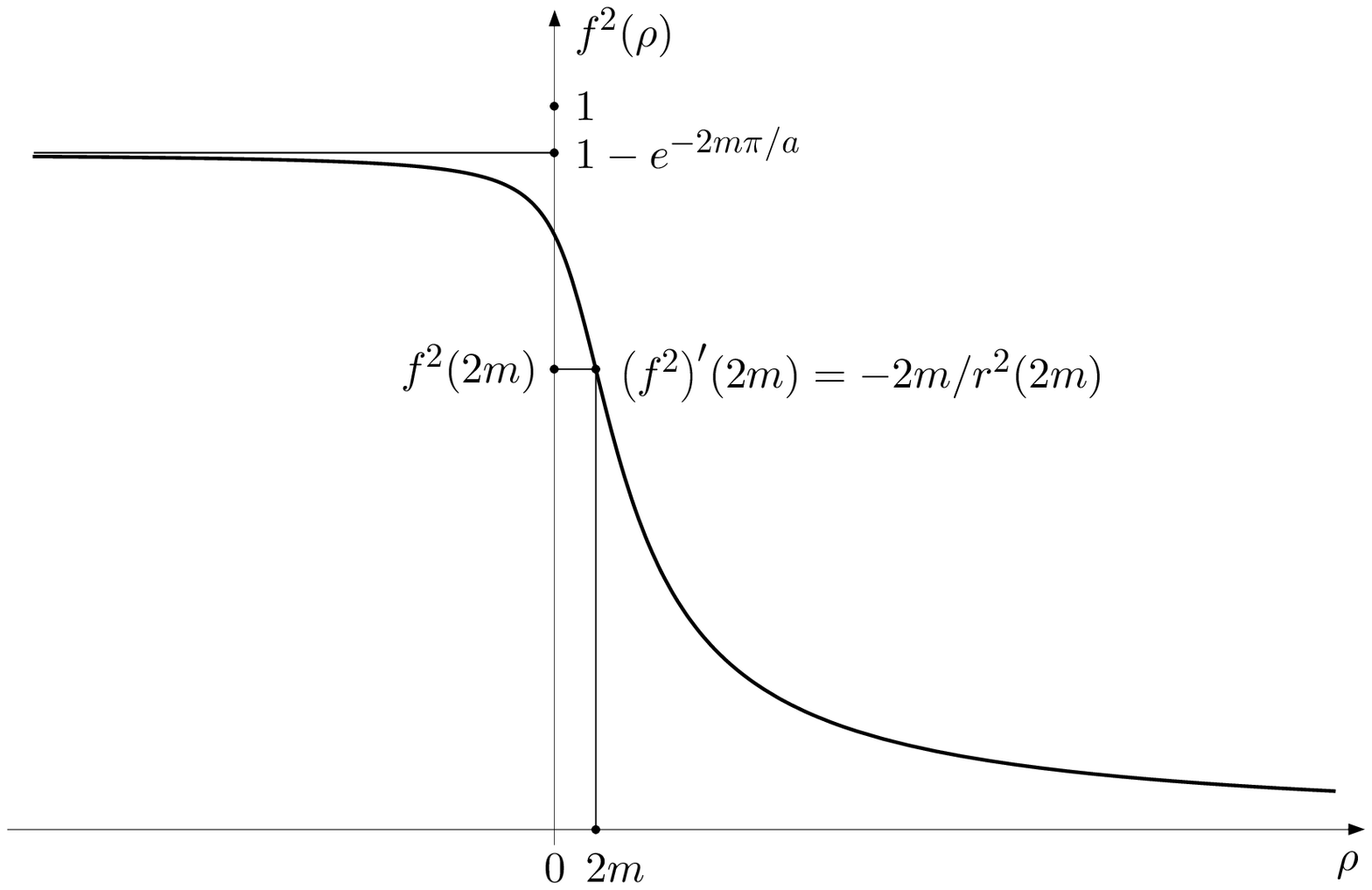}
\caption{\label{fig1} Graphs of $r(\rho)$ and $f^2(\rho)$ for generic values of
the parameters $m$ and $n$ ($0 \leq m < n$ and
$a \coloneqq \sqrt{n^2 - m^2})$.}
\end{figure}

As a replacement for the Schwarzschild blackhole model of a gravitating point
particle I derived in~\cite{elli2} a geodesically complete, spherically
symmetric, static solution of Eq.~(\ref{eqn7}), which I described as a
`drainhole' with `ether' flowing through it.  The metric is
\begin{equation}
d\tau^2 =
 [1 - f^2(\rho)] \, dt^2 - \frac{1}{c^2} \, [1 - f^2(\rho)]^{-1} \, d\rho^2
                                - \frac{1}{c^2} \, r^2(\rho) \, d\Omega^2 \, ,
\label{eqn8}
\end{equation} 
where
\begin{equation}
f^2(\rho) = 1 - e^{-(2 \, m/n) \alpha(\rho)} \, ,
\label{eqn9}
\end{equation}
\begin{equation}
r(\rho) = \sqrt{(\rho - m)^2 + a^2} \, e^{(m/n) \alpha(\rho)} \, ,
\label{eqn10}
\end{equation}
\noindent and
\begin{equation}
\phi = \alpha(\rho)
     = \frac{n}{a}
       \left[\frac{\pi}{2}
             - \tan^{-1} \left(\frac{\rho - m}{a}\right)\right] \, ,
\label{eqn11}
\end{equation}
\noindent
with $a \coloneqq \sqrt{n^2 - m^2}$, the parameters $m$ and $n$ satisfying
$0 \leq m < n$.  (Note that now $\rho$ is a radial coordinate unrelated to the
density parameter $\rho$ appearing in Einstein's $T^{\alpha \beta}$.  Also, the
coordinate $\rho$ used here translates to $\rho + m$ in~\cite{elli2}.) As
$\rho \to \infty$, $r(\rho) \sim \rho$ and $f^2(\rho) \sim 2 m/\rho$, which
shows the manifold to be asymptotic as $\rho \to \infty$ to the Schwarzschild
exterior solution, which is Eq.~(\ref{eqn8}) with
$r(\rho) = \rho$ and $f^2(\rho) = 2 m/\rho$.
The shapes and linear asymptotes of $r$ and $f^2$ are shown in Fig.~\ref{fig1}.

The metric of Eq.~(\ref{eqn8}) takes the more transparent `ether-flow' form
\begin{equation}
d\tau^2 =
 d{\bar t}\,^2 - \frac{1}{c^2} \, [d\rho - f(\rho) \, c \, d{\bar t}\,]\,^2
                      - \frac{1}{c^2} \, r^2 (\rho) \, d\Omega^2
\label{eqn12}
\end{equation}
when the coordinate $t$ is replaced by
$\displaystyle \bar t \coloneqq t
 - (1/c) \int f(\rho)[1 - f^2(\rho)]^{-1} \, d\rho$.
Because $r(\rho)$ is defined for $-\infty < \rho < \infty$ and has a positive
minimum value, the spatial cross sections of constant $\bar t$, with metric
$d\rho^2 + r^2(\rho) \, d\Omega^2$, comprise topologically two (nonflat)
euclidean three-manifolds, each with a spherical region excised, surgically
joined to make a nonsimply connected three-manifold with a spherical hole (the
`drainhole') in it, a configuration often referred to as an `Einstein--Rosen
bridge'.  The vector field
$\partial_{\,\bar t} + f(\rho) \, c \, \partial_\rho$, with
$f(\rho) \coloneqq -\sqrt{f^2(\rho)}$, is the velocity field of a cloud of test
particles (the `ether' cloud) free-falling from rest at $\rho = \infty$,
down into and through the drainhole, and on to $\rho = -\infty$, gaining speed
all the way, but never reaching the speed of light.

One would like to find a solution of the new, improved field equations to
replace the Schwarzschild interior solution for a homogeneous spherical ball
$\cal B$ of incompressible fluid, matched as smoothly as possible at the
surface of $\cal B$ to a relevant piece of the drainhole solution.  This
appears not to be doable without resorting to numerical calculations.  It is
possible, however, to find in `closed form' a solution that matches up smoothly
to the Schwarzschild exterior solution, as I shall now show.

To begin, let us introduce a coframe system 
$\{\omega^0, \omega^1, \omega^2, \omega^3\}$
orthonormal with respect to the metric of Eqs.~(\ref{eqn8}) and (\ref{eqn12}),
viz.,
\begin{equation}
\omega^0 = c \, d\bar t \, , \quad
\omega^1 = d\rho - f(\rho) \, c \, d\bar t \, , \quad
\omega^2 = r(\rho) \, d\vartheta \, , \quad
\omega^3 = r(\rho) \, \sin(\vartheta) \, d\varphi \, ,
\label{eqn13}
\end{equation}
dual to the frame system $\{e_0, e_1, e_2, e_3\}$ given by
\begin{equation}
e_0 = \partial_{\,\bar t} + f(\rho) \, c \, \partial_\rho \, , \quad
e_1 = \partial_\rho \, , \quad
e_2 = \frac{1}{r(\rho)} \, \partial_\vartheta \, , \quad
e_3 = \frac{1}{r(\rho) \sin(\vartheta)} \, \partial_\varphi \, .
\label{eqn14}
\end{equation} 
In this system the nonvanishing components of the Ricci tensor are
\begin{align}
{\bm R}_{00}
 &= -\nabla^2 \left(\frac{f^2}{2}\right) - 2 \frac{r''}{r} f^2 \, , \quad
    {\bm R}_{01} = {\bm R}_{10} = -2 \frac{r''}{r} f \, ,
\label{eqn15} \\
{\bm R}_{11}
 &= \nabla^2 \left(\frac{f^2}{2}\right) - 2 \frac{r''}{r} \, , \quad \text{and} \quad
{\bm R}_{22}
 = {\bm R}_{33} = \frac{1 - \left[(1 - f^2)(r^2/2)'\right]'}{r^2} \, .
\label{eqn16}
\end{align}
where $\nabla^2 h \coloneqq (1/r^2)(r^2 h')'$ for functions $h(\rho)$.  If
$\phi = \alpha(\rho)$, the nonvanishing products $\phi_{.\alpha} \phi_{.\beta}$
are
\begin{equation}
\phi_{.0} \phi_{.0} = \alpha'^{\,2} f^2 \, , \quad
\phi_{.0} \phi_{.1} = \phi_{.1} \phi_{.0} = \alpha'^{\,2} f \, , \quad
 \text{and} \quad
\phi_{.1} \phi_{.1} = \alpha'^{\,2} \, .
\label{eqn17}
\end{equation}
In terms of these the field equations~(\ref{eqn7}) reduce to
\begin{align}
-\nabla^2 \left(\frac{f^2}{2}\right) - 2 \frac{r''}{r} f^2
 &= \frac{4 \pi \kappa}{c^2} \mu - 2 \alpha'^{\,2} f^2 \, ,
\label{eqn18} \\
-2 \frac{r''}{r} f
 &= -2 \alpha'^{\,2} f \, ,
\label{eqn19} \\
\nabla^2 \left(\frac{f^2}{2}\right) - 2 \frac{r''}{r}
 &= -\frac{4 \pi \kappa}{c^2} \mu - 2 \alpha'^{\,2} \, ,
\label{eqn20} \\
\intertext{and}
\frac{1 - \left[(1 - f^2)(r^2/2)'\right]'}{r^2}
 &= -\frac{4 \pi \kappa}{c^2} \mu \, .
\label{eqn21}
\end{align}
An equivalent set is
\begin{align}
\left[(r^2)'(1 - f^2)\right]' &= 2 + \frac{8 \pi \kappa}{c^2} \mu \, r^2 \, ,
\label{eqn22} \\
\left[r^2 (1 - f^2)'\right]' 
 &= \frac{8 \pi \kappa}{c^2} \mu \, r^2 \, ,
\label{eqn23} \\
\intertext{and}
\frac{r''}{r}
 &= \alpha'^{\,2} \, .
\label{eqn24}
\end{align}

The drainhole metric is obtained by solving these equations with $\mu = 0$ and
rejecting solutions with singularities.  The Schwarzschild exterior metric for
the spherical ball $\cal B$ of radius $R$ is a solution of these equations with
$\mu = 0$ and $\alpha' = 0$.  In the `ether-flow' form of Eq.~(\ref{eqn12}) it
reads
\begin{equation}
d\tau^2 =
 d{\bar t}\,^2
  - \frac{1}{c^2} \left[d\rho + \sqrt{\frac{2 m}{\rho}} \, c \, d{\bar t}\right]^2
  - \frac{1}{c^2} \, \rho^2 \, d\Omega^2 \quad (\text{with } \rho \geq R) \, .
\label{eqn25}
\end{equation}
To search for a solution that extends it to the interior of $\cal B$ let us
take $\mu$ to be a nonzero constant and keep $\alpha' = 0$.  Then
Eq.~(\ref{eqn24}) implies that $r$ is linear, thus that
$r(\rho) = \lambda (\rho - \rho_0)$,
where $\rho_0$ is the value of $\rho$ at the center of $\cal B$ and $\lambda$
($= r'$) is to be determined later.  With $r$ so specified Eq.~(\ref{eqn23})
can be integrated once to yield
\begin{equation}
\left(1 - f^2\right)'
 = \frac{B}{r^2} + \frac{8 \pi \kappa \mu}{3 c^2 \lambda} \, r \, ,
\label{eqn26}
\end{equation}
and again to produce
\begin{equation}
1 - f^2 = A - \frac{B}{\lambda r}
            + \frac{4 \pi \kappa \mu}{3 c^2 \lambda^2} \, r^2 \, .
\label{eqn27}
\end{equation}
To avoid a singularity at the center of the ball, $B$ must be 0, and then
$A$ is fixed by substitution of $1 - f^2$ into Eq.~(\ref{eqn22}), the result
being that $A = 1/\lambda^2$, so that
\begin{equation}
1 - f^2
 = \frac{1}{\lambda^2} \left(1 + \frac{4 \pi \kappa \mu}{3 c^2} \, r^2\right).
\label{eqn28}
\end{equation}
Examining the metric $d\rho^2 + r^2(\rho) \, d\Omega^2$ of cross sections of
constant $\bar t$ one sees that a spherical surface defined by
$\rho = \text{constant} \leq R$ has geodesic radius $\rho$ (measured from
$\rho_0$, where $r = 0$) and areal radius $r(\rho)$ (area $= 4 \pi r^2(\rho$)).
The metric of a constant $\bar t$ cross section of the Schwarzschild exterior
solution is $d\rho^2 + \rho^2 \, d\Omega^2$, so a spherical surface defined by
$\rho = \text{constant} \geq R$ has areal radius $\rho$.  For the areal radii
to coincide at the surface of $\cal B$ it is necessary and sufficient that
$r(\rho)$ should be $R$ when $\rho = R$, that is,
$r(R) = \lambda (R - \rho_0) = R$,
from which follows $\rho_0 = \frac{\lambda - 1}{\lambda} \, R$.

Calculating the volume $V$ of $\cal B$ one has
\begin{equation}
V = \int_{\rho_0}^R 4 \pi r^2(\rho) \, d\rho
  = 4 \pi \lambda^2 \int_{\rho_0}^R (\rho - \rho_0)^2 \, d\rho
  = \frac{4 \pi}{3 \lambda} \, R^3 \, .
\label{eqn29}
\end{equation} 
Let $M$ be the {\it active} gravitational mass of $\cal B$ and let
$m \coloneqq \kappa M/c^2$ (i. e., $m = M$ in geometric units).  Then
$M = \mu V$, from which follows $4 \pi \mu = 3 \lambda M/R^3$, thus that
\begin{equation}
1 - f^2
 = \frac{1}{\lambda^2}
   \left(1 + \frac{\lambda \kappa M}{c^2 R} \,\frac{r^2}{R^2}\right)
 = \frac{1}{\lambda^2}
   \left(1 + \frac{\lambda m}{R} \, \frac{r^2}{R^2}\right) \, .
\label{eqn30}
\end{equation}
There remains only to determine $\lambda$ by application of the matching of the
interior solution's $f^2(\rho)$ to the exterior solution's $2 m/\rho$ at the
surface of $\cal B$, where $r(\rho) = \rho = R$.  In light of Eq.~(\ref{eqn30})
this comes down to the equation
\begin{equation}
\frac{1}{\lambda^2} \left(1 + \frac{\lambda m}{R}\right)
 = 1 - f^2(R)
 = 1 - \frac{2 m}{R} \, ,
\label{eqn31}
\end{equation}
which is equivalent to
\begin{equation}
\left(1 - \frac{2 m}{R}\right) \lambda^2 - \frac{m}{R} \lambda - 1 = 0 \, .
\label{eqn32}
\end{equation}
Presuming that $R > 2 m$ (the areal radius of the blackhole event horizon that
$\cal B$ would have if $\cal B$ were small enough to have an event horizon),
and that $r' = \lambda > 0$, one has from Eq.~(\ref{eqn31}) that 
\begin{equation}
\lambda^2 = \frac{1 + \lambda m/R}{1 - 2 m/R} > 1 \, ,
\label{eqn33}
\end{equation}
thus that the areal radius $r(\rho) = \lambda (\rho - \rho_0)$ increases from 0
to $R$ more rapidly than the geodesic radius $\rho$ increases from $\rho_0$ to
$R$, and from Eq.~(\ref{eqn32}) that
\begin{equation}
\lambda = \frac{m + \sqrt{m^2 + 4 R (R - 2 m)}}{2 (R - 2 m)} \, .
\label{eqn34}
\end{equation}

Collecting these results, one has as the interior metric of $\cal B$ derived
from the new, improved field equations the following replacement for the
Schwarzschild interior metric of Eq.~(\ref{eqn1}):
\begin{equation}
d\tau^2
 = \frac{1 - 2 m/R}{1 + \lambda m/R}
   \left[1 + \frac{\lambda m}{R} \, \frac{r^2(\rho)}{R^2}\right] \, dt^2
   - \frac{1}{c^2} \frac{1 + \lambda m/R}{1 - 2m/R}
     \left[1 + \frac{\lambda m}{R} \,
               \frac{r^2(\rho)}{R^2}\right]^{-1} \! d\rho^2
   - \frac{1}{c^2} \, r^2(\rho) \, d\Omega^2 \, ,
\label{eqn35}
\end{equation}
where $r(\rho) = \lambda (\rho - \rho_0)$,
$\rho_0 = \frac{\lambda - 1}{\lambda} R$, and $\lambda$ is given by
Eq.~(\ref{eqn34}).  The `ether-flow' version is
\begin{equation}
d\tau^2
 = d{\bar t}\,^2
   - \frac{1}{c^2} \left[d\rho
           + \sqrt{1 - \frac{1 - 2 m/R}{1 + \lambda m/R}
                       \left(1 + \frac{\lambda m}{R} \,
                                 \frac{r^2(\rho)}{R^2}\right)} \,
                                 c \, d\bar t\right]^2
   - \frac{1}{c^2} \, r^2(\rho) \, d\Omega^2 \, .
\label{eqn36}
\end{equation}

One can show (cf.~\cite{elli2}, Eq.~(53)), that the radial equation of motion
for a test particle following a timelike geodesic of the metric of
Eq.~(\ref{eqn12}), parametrized by proper time $\tau$, is
\begin{equation}
\ddot \rho
 = \left(\frac{f^2}{2}\right)' c^2
   + \frac12 [(r^2)' (1 - f^2) - r^2 (1 - f^2)'] \, {\dot \Omega}^2 \, .
\label{eqn37} 
\end{equation}
For the Schwarzschild exterior metric this becomes
\begin{equation}
\ddot \rho = -\frac{m c^2}{\rho^2} + (\rho - 3 m) \, {\dot \Omega}^2
           = -\frac{\kappa M}{\rho^2} + (\rho - 3 m) \, {\dot \Omega}^2 \, ,
\label{eqn38}
\end{equation}
and for the new interior metric
\begin{equation}
\ddot \rho
 = -\frac{m c^2}{R^2} \frac{r(\rho)}{R}
    + \frac{1}{\lambda} \, r(\rho) \, {\dot \Omega}^2
 = -\frac{\kappa M}{R^2} \frac{r(\rho)}{R}
    + \frac{1}{\lambda} \, r(\rho) \, {\dot \Omega}^2 \, .
\label{eqn39}
\end{equation}
On the boundary of $\cal B$ these collapse to
\begin{equation}
\ddot \rho = -\frac{m c^2}{R^2} + (R - 3 m) \, {\dot \Omega}^2
           = -\frac{\kappa M}{R^2} + (R - 3 m) \, {\dot \Omega}^2
\label{eqn40}
\end{equation}
and
\begin{equation}
\ddot \rho
 = -\frac{m c^2}{R^2} + \frac{R}{\lambda} \, {\dot \Omega}^2
 = -\frac{\kappa M}{R^2} + \frac{R}{\lambda} \, {\dot \Omega}^2 \, ,
\label{eqn41}
\end{equation}
which shows continuity at the boundary of the gravitational component of the
radial acceleration, but discontinuity of the centrifugal component, inasmuch
as $R/\lambda \neq R - 3 m$.  At the center of $\cal B$, according to
Eq.~(\ref{eqn39}), both components vanish so $\ddot \rho = 0$, as would be
expected from experience with newtonian gravitational theory.

With the interior metric of Eqs.~(\ref{eqn35})~and~(\ref{eqn36}) in hand, one
can study its geodesics in depth.  In particular, one can investigate geodesics
connecting two points A and B on the surface of $\cal B$ and compute the
distances traveled and times of flight of photons and neutrinos passing from A
to B (perhaps through a tunnel, to exclude nongravitational interactions with
the matter of $\cal B$).  Such an investigation is carried out in \cite{elli3},
with application of the results to the measurements described in \cite{icarus}
and \cite{opera2} of the flight times of neutrinos from a source point A at
CERN's European Laboratory for Particle Physics to points B in the OPERA and
ICARUS detectors at LNGS (Laboratori Nazionale del Gran Sasso). 

\eject

\begin{center} \rule{3.5in}{.01in} \end{center}

\noindent
Homepage: \url{http://euclid.colorado.edu/~ellis}
\hfill
Electronic mail: {ellis@euclid.colorado.edu}

\end{document}